\documentclass[8pt, a4paper, twocolumn]{article}
\usepackage{amsmath}
\usepackage{amssymb}
\usepackage{cite}

\title{ Swampland conjectures and single field inflation in non-standard cosmological scenarios }

\author{Oem Trivedi \footnote{oem.t@ahduni.edu.in}}
\date{%
	School of Arts and Sciences, Ahmedabad University,Ahmedabad 380009,India\\%
	\today
}
\begin{document}
	
	\maketitle

\begin{abstract}
	Swampland conjectures have attracted quite some interest in the cosmological community. They have been shown to have wide ranging implications , like constraints on inflationary models, primordial black holes, dark energy to name a few. Particularly, their implications on single field inflationary models in general relativity based cosmology has gathered huge attention. Swampland conjectures in their usual form have been shown to be incompatible with these kind of single field models, or have been shown to induce severe fine tuning in these inflationary models for them to be consistent with the conjectures. In this work, we show that a large class of single field inflationary models can in fact bypass the problems faced by inflationary paradigms in GR Based cosmology. We use the exact solution approach to inflation for the same purpose and show how string theoretic motivations of the swampland conjectures can be in perfect symphony with various single field inflationary models in non-standard cosmological scenarios.
	
\end{abstract}

\section{Introduction }

The idea of cosmic inflation has achieved a tremendous amount of success in describing the early universe. inflation initially captured attention with it's solutions to the fine tuning problems of standard big bang cosmology \cite{guth1981inflationary,sato1981first,starobinskii1979spectrum,linde1995quantum,baumann2009tasi}. The optimism regarding inflationary cosmology kept on increasing as it's various observational predictions about the early universe were shown to be very precise in accordance with various satellite experiments, a trend which continued even in the Planck 18' results \cite{aghanim2018planck,akrami2018planck}.A variety of inflationary models are supported by the observational data ,and amongst the most famous of these paradigms are  inflationary models. In these models, inflation is driven by a single scalar field, popularly referred to as the "inflaton" \cite{baumann2009tasi}. While the usual paradigm of single field models considers only a real scalar field which dominates the energy density of the universe in early times, there are some other regimes where non standard considerations are taken about the field nature or the way it interacts \cite{berera1995warm,golovnev2008vector}. But still the conventional single field models (which some would like to call "supercooled inflationary models" \cite{berera1995warm}) are very well supported observationally and are very prevalent in theoretical studies too. 
String theory has attracted a resounding amount of attention in the cosmological community ever since it's inception \cite{mcallister2008string,baumann2015inflation,gasperini2007elements,heckman2019f,heckman2019pixelated}.As one of the most well studied theories which promise to describe " quantum gravity",one may expect string theory to appropriately describe the workings of the early universe. Indeed it has been found that the "string landscape"(the set of all possible vacua admitted by the theory), is quite large going to $ \mathcal{O} (10^{500}) $. Hence one can expect that string theory can be a good candidate for a " theory of everything" for our universe.But then the question of how one could distinguish between effective field theories which are consistent and those which are non consistent with string theory. To this end, Vafa in \cite{vafa2005string} coined the term "swampland", referring to the class of effective field theories incompatible with a self consistent theory of quantum gravity. Further, to distinguish between theories in the swampland and those compatible with self consistent quantum gravity, Vafa and his co authors proposed a number of "swampland conjectures" \cite{ooguri2007geometry,ooguri2016non,mcnamara2019cobordism,bedroya2019trans}. The prominent swampland conjectures which gathered immediate interest in the context of inflationary cosmology were :
$ 1 $ : Swampland distance conjecture (SDC) : This conjecture limits the field space of validity of any effective field theory \cite{ooguri2007geometry}. This sets a maximum range traversable by the scalar fields in an EFT as \begin{equation}
\Delta \phi \leq d \sim \mathcal{O} (1)
\end{equation} 
where we are working in the Planck Units $ m_{p} = 1 $ where $ m_{p} $ is the reduced Planck's constant, d is some constant of $ \mathcal{O} (1) $ , and $\phi$ is the scalar field of the EFT. 
$ 2 $ Swampland De Sitter conjecture (dS conjecture) : This conjecture states that it is not possible to create dS Vacua in String theory \cite{ooguri2016non}. The conjecture is a result of the observation that it has been very hard to generate dS Vacua in String theory \cite{dasgupta2019sitter,danielsson2018if}( While it has been shown that creating dS Vacua in String theory is possible in some schemes ,like the KKLT construction \cite{kachru2003sitter}). The Conjecture sets a lower bound on the gradient of scalar potentials in an EFT , \begin{equation}
\frac{| V^{\prime} |}{V} \geq c \sim \mathcal{O} (1)
\end{equation} 
where c is some constant of $ \mathcal{O} (1) $ , and V is the scalar field potential. Another " refined " form of the swampland De Sitter conjecutre places constraints on the Hessian of the scalar potential (a finding which first appeared in \cite{garg2018bounds} and later in \cite{ooguri2019distance} ). Expressed in $ m_{p} = 1 $ units, it reads \begin{equation}
\frac{V^{\prime \prime}}{V} \leq - c^{\prime} \sim \mathcal{O} (1)
\end{equation}
where $ c^{\prime} $ is again some constant of Order 1. 
These criterion have quite strong implications across different fields of physics. It's pretty revealing consequences on the standard model were discussed in \cite{march2020qcd}.Here it was shown that although the SM is compatible with the conjectures so far, changing the value of various a-priori free parameters, such as masses of quarks and the Higgs boson, would plunge the SM into the swamp. The most exciting consequence of this might be to tie certain unexplained features of the SM, such as the notoriously problematic lightness of the Higgs to quantum gravity.These conjectures also have a very interesting say on the nature of dark energy \cite{agrawal2018cosmological,odintsov2019finite,chiang2019does} and dark matter \cite{kawasaki2018primordial,agrawal2019h_0}.Further,it has been shown that  inflation in a universe described by general relativity based cosmology is incompatible with the swampland criterion, in a particular the dS condition, for a general class of potentials\cite{kinney2019zoo}. The disagreement can be seen in essence through the slow roll parameters for  GR Based inflation. The primary slow roll parameters of usual  inflation are the famous $\epsilon$ and $\eta$ parameters, and it is well known that for inflation to occur both of them should be very less than unity \cite{baumann2009tasi}. Defining $\eta$ and $\epsilon$, in Planck Units the condition is given as \begin{equation}
\epsilon = \frac{1}{2} \left( \frac{V^{\prime}}{V} \right)^{2}
\end{equation}
\begin{equation}
\eta = \frac{V^{\prime \prime}}{V}
\end{equation}	
It is immediately seen from the definitions of the $\epsilon$ parameter (4) , and the dS Conjecture (2) , that these  conditions are in conflict with each other. Furthermore, a new swampland criteria by the name of " Trans Planckian Censorship Conjecture(TCC)" \cite{bedroya2019trans} , implies that single field GR based inflationary models would have to be severely fine tuned for them to not be in the swampland \cite{bedroya2020trans}. A lot of work has been done to further understand the relations of this conjecture with inflation \cite{dhuria2019trans,brahma2020trans,mizuno2020universal,cai2020mass,kamali2020relaxing}.While these works seem to show that inflationary cosmology and swampland criterion are in direct logger heads, hence a lot of research has gone into understanding whether there could be ways to reduce, if not fully resolve, the conflict between the conjecture and inflationary models.  A lot of work has gone into resolving the conflict between single field inflation and the conjectures. In \cite{geng2020potential} it was found that in a certain regime on the parameter space the back reaction from emergent modes due to distance conjecture can render slow roll consistent with de-sitter conjecture and its refinement.In \cite{scalisi2020inflation} the implications of SDC on inflation were discussed for the tower of states having spin larger than 1 and it was shown that the only way to avoid conflict with single field inflation was to have a cut off spin value.\cite{scalisi2019swampland} also discussed about the implications on inflationary $\alpha$ attractors. \cite{brahma2020swampland} discussed the general properties required for avoiding the swampland in single-field inflation while relationships between different swampland conjectures in an inflating space-time were pointed out in \cite{brahma2019relating}. While in \cite{ashoorioon2019rescuing}, it was shown how one can salvage single field inflation given the swampland conjectures using excited initial states for tensor perturbations. While this great amount of literature has produced some really insightful and encouraging results, Fine Tuning of key physical parameters still remains a big issue for single field inflation in a GR based cosmology given the swampland conjectures. 
However, inflation stands on more amicable standings with the conjectures in certain other different regimes of inflation besides single field GR based models. Multi field inflation has been shown to be consistent with the swampland conjectures (3) and (4). \cite{achucarro2019string,bravo2020tip,damian2019two}.   
The paradigm of Warm inflation, even for single fieldmodels, is consistent with the swampland conjectures \cite{berera2019trans,motaharfar2019warm,das2019warm,das2020swampland} .

Alongside the usual GR based cosmology, an enticing possibility for inflationary cosmology in recent decades has been the study of inflation in various non-standard cosmological scenarios. These include the Randall-Sundrum and other braneworld scenarios from  string theory itself \cite{randall1999alternative,randall1999large,gogberashvili2002hierarchy}, cosmologies due to modified gravitational scenarios  \cite{cai2008corrected,tsallis2013black,gomez2011standard} , loop quantum cosmologies \cite{ashtekar2006quantum,ashtekar2006quantum1,ashtekar2006quantum2} amongst others. The paradigm of single scalar field inflation has been studied substantially in all the above mentioned regimes as well. braneworld inflation has been studied both in the supercooled and warm regimes \cite{setare2014constructing,hawkins2001inflation,choudhury2003hybrid,kallosh2003p,del2007warm,cid2007warm,matsui2019eternal}, loop quantum inflation was explored in \cite{bhardwaj2019inflation} while \cite{del2012approach} studied the modified Chern-Simons cosmology. It has also been shown that single field inflation in non-standard cosmological scenarios can be quite easily consistent with swampland conjectures \cite{kamali2020warm,lin2019chaotic,blumenhagen2017swampland,odintsov2020swampland,bernardo2020trans,adhikari2020power}. An important point to note here is that the swampland conjectures have widely been intended to be criterion applicable for 4-D low energy EFT's with canonically normalized minimally coupled fields (although recently there has been a lot of work which has considered the swampland bounds for inflation in non canonical regimes too \cite{seo2019sitter,mohammadi2020warm,solomon2020non,trivedi2021rejuvenating,lin2021generalizing}). The minimal coupling requirement can be easily understood to be fulfilled in the non-GR based scenarios, as its clear with the inflaton field being in a standard Klein-Gordon form for minimally coupled fields. The requirement that the regime in which the swampland conjectures are being studied be effective 4-D scenarios can be satisfied even in non-GR setups where gravity might be considered to have a higher dimensional origin, like in RS-II braneworld inflationary scenarios \cite{lin2019chaotic,adhikari2020power,kamali2020warm} (which lies in the scope of the analysis we are going to be describing in this work). The RS-II scenario is indeed a good example to make the higher dimensional gravity point clear, as the inflationary regime which comes about by considering the RS-II realization is one in which inflation is still considered in a 4-D world with its space dimensions lying on a 3 Brane, which is embedded in a five dimensional bulk. In a similar vein, Chern-Simons models(where the gravitational theory is again considered to have 5-D origin but whose inflationary regime was studied in an effective 4-D regime just like RS-II scenario in \cite{del2012approach}) and the alike scenarios possessing a higher dimensional grounding which are in the scope of our analysis below are indeed fit for use with the swampland conjectures.  single field supercooled inflation in non GR based cosmological scenarios being in extremely good agreements with swampland conjectures might suggest that a self consistent quantum gravitational theory would point towards the early universe having a different cosmological setup than general relativity (considering that the swampland conjectures indeed describe conditions a self consistent theory of gravity would uphold).\\
In this paper, we show that a wide variety of single scalar field inflationary models in non-standard cosmological scenarios  can satisfy the swampland criteria. In Section 2 , we will briefly describe the exact solution approach to inflation \cite{del2012approach} which we will use in the whole of our paper. In Section 3 , use the exact solution approach and show how single field inflation can satisfy swampland constraints for a general class of non-standard cosmological scenarios. We conclude our work in Section 4 with comments on the scope of applicability of our analysis. 

\section{The exact solution approach to inflation }

The exact solution approach for single field(Supercooled) inflationary models was introduced by Del Campo in \cite{del2012approach}. The crucial point of note is that a generalized Friedmann equation of the form, \begin{equation}
F(H) = \frac{8 \pi}{3 m_{p}^{2}} \rho_{\phi}
\end{equation} 
where we have written the Friedmann equation in the units used in \cite{del2012approach} , $ c =  \hbar = 1 $ and $\rho_{\phi}$ is the energy density of the inflaton field and $ F(H) $ is a general function of the Hubble parameter $ H = H(\phi) $ ( $\phi$ dependent Hubble parameters were famously first considered in the Hamilton Jacobi approach to inflation of Kinney \cite{kinney1997hamilton} ) . This equation is quite general in the sense that one can arrive at the Friedmann equation of different cosmologies for different $ F(H) $. For Instance, the braneworld scenario Friedmann equation is attained for \cite{shiromizu2000einstein} for \begin{equation}
F(H) = \left( \frac{8 \pi \lambda}{3 m_{p}^{2}} \right) \bigg[ \sqrt{1 + \left( \frac{3 m_{p}^{2}}{4 \pi \lambda} \right) H^{2} } - 1 \bigg]
\end{equation} The Chern-Simons model \cite{gomez2011standard} can be attained for \begin{equation}
F(H) = H^{2} - \alpha H^{4}
\end{equation}
Various exotic gravitational models can also be encapsulted in (5), like for example a modified area-entropy cosmology where the entropy is of the Tsallis type \cite{sheykhi2018modified} \begin{equation}
F(H) = H^{2(2-\kappa)}
\end{equation}
where $\kappa$ is the Tsallis parameter and similarly, one can even get loop quantum cosmology \cite{ashtekar2006quantum2} with (5).

As is well known  during slow roll inflation, the field velocity $ \dot{\phi} = \frac{d \phi}{dt} $ is very less than the scalar potential $ V(\phi) $,  \begin{equation}
\dot{\phi}^{2}  << V(\phi)
\end{equation}  
which allows us to write $ \rho_{\phi} \approx V(\phi) $. The equation of motion which the scalar field satisfies is the usual Klein Gordon equation, \begin{equation}
\ddot{\phi} + 3 H \dot{\phi} + V^{\prime}(\phi) = 0
\end{equation}
where $ \ddot{\phi}  = \frac{d^{2} \phi}{dt^{2}} $ 

based on these basics ideas, the field velocity is given by \begin{equation}
\dot{\phi} = - \frac{m_{p}^{2}}{8 \pi} \frac{F_{_{\prime} H} H^{\prime}}{H}
\end{equation} one can further find the $\epsilon$ and $\eta$ parameters to be, \begin{equation}
\epsilon = \frac{m_{p}^{2}}{8 \pi} \frac{F_{_{\prime} H}}{H} \left( \frac{H^{\prime}}{H} \right)^{2}
\end{equation}
\begin{equation}
\eta = \frac{m_{p}^{2}}{8 \pi} \frac{F_{_{\prime} H}}{H} \frac{H^{\prime \prime}}{H}
\end{equation}
where $ F_{_{\prime} H} = \frac{\partial F}{\partial H} $ and $ H^{\prime} = \frac{dH}{d\phi} $  The Number of e-folds is given by, \begin{equation}
N = \frac{8 \pi}{m_{p}^{2}} \int_{\phi_{e}}^{\phi} \frac{H^{2}}{H^{\prime} F_{_{\prime} H}} d \phi
\end{equation}
One can further define the horizon flow equations and comment on the attractor nature of the inflationary model in question using this approach, as sh  own in \cite{del2012approach}. 

The spectra of scalar and tensor perturbations are of special interest for any inflationary model as they allow the model to be tested using  observational data \cite{baumann2009tasi}. Using the definition of the slow roll parameters and the usual formulation of the amplitude squared scalar perturbations at horizon exit $ k = aH $, \begin{equation}
P_{\mathcal{R}} (k) = \left( \frac{H}{| \dot{\phi} |} \right)^{2} \left( \frac{H}{2 \pi} \right)_{a H = k}
\end{equation}
we get the scalar spectral index as , \begin{equation}
n_{s} =  1 + \frac{d P_{\mathcal{R}} (k)}{d ln k} = 1 + 2 \eta  -2 \epsilon \left( 3 - H \frac{F_{{\prime } H H} }{F_{_{\prime}H}} \right)
\end{equation}
where $  F_{_{\prime} H H} =  \frac{\partial^{2} F }{\partial H^{2}}  $. Similarly the squared amplitude of tensor perturbations is given by , \begin{equation}
P_{\mathcal{T}} (k) = \frac{16 \pi}{m_{p}^{2}} \left( \frac{H}{2 \pi} \right)_{a H = k}^{2}
\end{equation}
And hence the tensor spectral index is given as, \begin{equation}
n_{T} = -2 \epsilon
\end{equation}
The tensor to scalar ratio is hence given by, \begin{equation}
r = 2 \frac{F_{_{\prime} H}}{H} \epsilon
\end{equation}
While one can also get the running of both scalar and tensor spectral index, we will not pursue it at this time. This completes a brief discussion of the fundamentals of the  which will prove to be crucial for our analysis of the swampland issues of single field inflation. It is important to note here that this approach towards inflation takes into account all kinds of modifications that one would observe in various important physical quantities for inflation like the slow roll parameters, scalar and tensor spectral index, tensor-to-scalar ratio etc. in modified cosmologies. Whether the modifications come through additional parameters of the cosmology like the brane tension for RS-II braneworld inflation or whether these are extra changes which these quantities get through some free parameters for some particular model in these cosmologies like in the consideration of some specific potential (or equivalently, in the consideration of some specific ansatz for the Hubble parameter), the  takes into account all such changes and gives us an appropriately modified treatment of inflation for different cosmologies. The reader can refer to \cite{del2012approach} for having a look at a more detailed description of this approach and in particular how it can successfully demonstrate inflation in paradigms like RS-II braneworld and the Chern-Simons cosmology(giving all the needed expressions for the inflationary setup, right from the slow roll parameters to all the perturbation parameters too), which were discussed in length using this approach in that work.
Now, we have cleared the basics of the exact solution approach which we will be using in the next section.
\section{Consistency of single field models in non-standard cosmologies }
We will use the planck units $ m_{p} = \sqrt{1/8 \pi G} = 1 $. So our generalized Friedmann equation (5) takes the form, \begin{equation}
F(H) = \frac{\rho_{\phi}}{3}
\end{equation}
Which during inflation($ \dot{\phi}^{2} << V(\phi) $ ) becomes , \begin{equation}
F(H) = \frac{V}{3}
\end{equation}
This allows us to write the Gradient of the derivative of the scalar potential with respect to $\phi$ as , \begin{equation}
F_{_{\prime} H} H^{\prime} = \frac{V^{\prime} }{3}
\end{equation}
This allows us to write the ratio \begin{equation}
\frac{| V^{\prime} |}{V} = \frac{F_{_{\prime} H} H^{\prime}}{F}
\end{equation}
For now, let's call the above ratio k $$ k = \frac{V^{\prime}}{V} = \frac{F_{_{\prime} H} H^{\prime}}{F}
$$  Now, the $\epsilon$ parameter in $ m_{p} = 1 $ units is given by, \begin{equation}
\epsilon = \frac{F_{_{\prime} H} }{H} \left( \frac{H^{\prime}}{H} \right)^{2}
\end{equation}
Now, $\epsilon$ can be written in terms of k as , \begin{equation}
\epsilon = \frac{F H^{\prime} k }{H^{3} }
\end{equation}

Now, one of the central points of disagreement between single field inflationary models in GR based cosmologies and the swampland conjectures is the logger heads the smallness of $\epsilon$ parameter condition, $ \epsilon << 1 $ , is with the swampland dS conjecture (2). This conflict is , however, not evident in various investigations of inflation in non-standard cosmological scenarios  \cite{kamali2020warm,lin2019chaotic,blumenhagen2017swampland}.
Keeping that in mind, if we now write the dS conjecture in a non-standard  cosmology given by a Friedmann equation of the form (21) in terms of the ratio k, we have the dS criterion as \begin{equation}
k \geq c \sim \mathcal{O} (1)
\end{equation} 
This tells us that in order to satisfy the De Sitter conjecture in a non-standard cosmological scenario we need , \begin{equation}
\frac{H^3}{F H^{\prime}} >> \mathcal{O} (1)
\end{equation}
In which case $ \epsilon $ still remains considerably below unity even if we consider the dS conjecutre. 
Hence, a large class of cosmologies which are within the scope of the generalized Friedmann equation (21) can easily be overcome $\epsilon$ parameter issue with dS conjecture. For GR based cosmology, $ F(H) = H^{2} $, one recovers the conflict between the concerned conjecture and single field inflation as the exact solution approach reduces to Kinney's Hamilton-Jacobi Approach in that limit \cite{kinney1997hamilton}. 

Another point of conflict with the conjectures and single field inflationary models in GR based cosmologies is the issue of insufficient e-fold number \cite{kinney2019zoo}. It can be shown that the number of e-folds for single field GR based inflationary models can be given by, \cite{lin2019chaotic} , \begin{equation}
N \approx \frac{V \Delta \phi}{V^{\prime}}
\end{equation} 
If we move in accordance with the swampland conjectures, then $ N < \mathcal{O} (1) $ . This is a serious problem , as the most recent data on inflation \cite{akrami2018planck} requires around 50 to 60 Number of e-folds. It is again clearly evident that the swampland conjectures are in conflict with single field inflationary models in GR based cosmology. But once again, it has been shown that this problem does not arise in non-standard cosmological scenarios \cite{kamali2020warm,lin2019chaotic,blumenhagen2017swampland} . Now, we will look at the same problem within the view of the exact solution approach and show that this issue does not arise in a large class of non-standard cosmological scenarios. 

The e-fold Number for inflation \cite{del2012approach} expressed in $ m_{p} = 1 $ units is  , \begin{equation}
N = \int_{\phi_{e}}^{\phi} \frac{H^{2}}{F_{_{\prime} H} H^{\prime}} d \phi
\end{equation}
The above expression for the e-fold number can be rewritten as \begin{equation}
N = \int_{\phi_{e}}^{\phi} \frac{H^{2}}{F} \frac{1}{\frac{F_{_{\prime} H} H^{\prime}}{F}} d\phi
\end{equation}
Thus, e-fold number can be approximately given as , \begin{equation}
N \approx \frac{H^{2}}{F} \frac{\Delta \phi}{\frac{F_{_{\prime} H} H^{\prime}}{F}} =  \frac{H^{2}}{F} \frac{\Delta \phi}{k}
\end{equation}
Now, in accordance to the distance conjecture and  dS conjecture , $ \Delta \phi \leq \mathcal{O} (1) $ , $ k \geq 1 $ respectively. We see that in contrast to single field inflation in GR based cosmology, the e-fold number for single field models in a non-standard cosmological scenario can be high enough for sufficient inflation to occur. Again , in the GR Limit $ F= H^{2} $ the problem of insufficient e-fold number persists but in a generalized scenario it can be appropriately high.  
Another issue with single field inflationary models in GR based cosmology and swampland conjectures is the order of the parameter c in the definition of the dS conjecture. From string theory motivated constructions, c in (2) should be of the order of unity. While it was shown in \cite{kinney2019zoo} that in order for the conjectures and Single field models to be consistent with the value of the scalar spectral index of Planck 18' data \cite{akrami2018planck} , $ c \sim \mathcal{O} (0.1) $. This shows a stark contrast between the string theory motivations of the conjecture and their consistency with the observational data for inflationary cosmology. 

We now show that the string theoretic motivations of the swampland conjectures can indeed be consistent with the observational data for inflationary models in non-standard cosmological scenarios. From the equation (16) for $ n_{s} $ and the definitions of the $\epsilon$ and $\eta$ parameters (12-13) in $ m_{p} = 1 $ units  , we can express the scalar spectral index as, \begin{equation}
n_{s} = 1 + 2 \left( \frac{F_{_{\prime} H} }{H} \frac{H^{\prime \prime}}{H} \right) - 2 \frac{F_{_{\prime} H} }{H} \left( \frac{H^{\prime}}{H} \right)^{2} \big( 3 - H \frac{F_{_{\prime} H H}}{F_{_{\prime} H}}  \big)
\end{equation}

Now, the above equation can be rewritten as, \begin{equation}
n_{s} = 1 + 2 \frac{F_{_{\prime} H} H^{\prime}}{F} \frac{F H^{\prime \prime}}{H^{\prime} H^{2}} - 2 \frac{F_{_{\prime} H} H^{\prime}}{F} \frac{F H^{\prime}}{H^{3}} \bigg( 3 - H \frac{F_{_{\prime} H H}}{F_{_{\prime} H}}  \bigg)
\end{equation}
And finally after some rearrangement, one can arrive at the following relation \begin{equation}
\left(\frac{(1 - n_{s} ) H^{2}}{2F} \right) \bigg[ \frac{1}{\left( 3 - H \frac{F_{_{\prime} H H}}{F_{_{\prime} H}} \right) \frac{H^{\prime}}{H} - \frac{H^{\prime \prime}}{H^{\prime}}} \bigg] =  \frac{F_{_{\prime} H} H^{\prime}}{F} \geq c
\end{equation}  The above relation allows us to have an upper bound on c in terms of the observed value of the scalar spectral index in some cosmological scenario described by (21). We see that the required bound is qualitatively the same as the one needed for the $\epsilon$ parameter requirement (28), which is obvious as this problem found its roots in the $\epsilon$ parameter issue \cite{kinney2019zoo}.  Alongside this, we now make use of the refined dS conjecture  (3) to comment on the relationship between $\eta$ and 4 $ c^{\prime} $. Using (22) , we can write,\begin{equation}
F_{_{\prime} H H} {H^{\prime}}^{2} + F_{_{\prime} H} H^{\prime \prime} = \frac{V^{\prime \prime}}{3}
\end{equation} 
In the light of the above relation, one can express (3) in terms of $\eta$ (14) as \begin{equation}
\frac{1}{F_{_{\prime} H}} \bigg[ F_{_{\prime} H H} H^{\prime} + \eta \frac{F_{_{\prime} H} H^{2}}{H^{\prime}} \bigg] \leq - c^{\prime}
\end{equation}

Equations (36) and (38)  alongside (26) are especially important in understanding the distinction between implications of swampland conjectures for single field inflation in GR based cosmology and in more generalized scenarios. It was shown systematically in \cite{kinney2019zoo} how (2) and (3) lead the $\epsilon$ and $\eta$ parameters in single field GR based inflationary models to be constrained as $$ \epsilon \geq \frac{c^{2}}{2} $$ $$ \eta \leq -c^{\prime} $$  These relations lead in turn to \begin{equation}
1 - n_{s} \geq [3 c^{2} ; 2 c^{\prime}]
\end{equation}
Then, the authors in \cite{kinney2019zoo} used the data for the scalar spectral index and the tensor to scalar ratio \cite{akrami2018planck} to show that $ c \leq \mathcal{O} (0.1) $ and $ c^{\prime} \leq \mathcal{O} (0.01) $ . This led them to categorically demonstrate that there is a severe conflict with the string motivated definitions of c and $ c^{\prime} $ being order 1, and the observational data for inflation which constrain them to be smaller orders in order for single scalar field models in GR based cosmologies to be compatible with them. 

We now see from equations (38) , (36) and (26) that in a general non-standard cosmology described by (22) the strict relations of $\epsilon$ and $\eta$ to c and $ c^{\prime} $ do not hold. (26) shows that c can indeed be some order 1 term and still $\epsilon << 1 $. Similarly (36) shows that the scalar spectral index is not directly related to an upper bound for c , as found for single field models in \cite{kinney2019zoo}. Rather, in a general cosmology the free parameters of the models can be put in limits to fit up well with the observational data without violating in the Order 1 nature of c. Observational data also constrains $ \eta \leq 0 $ and this fact can also be easily satisfied alongside the swampland definition of $ c^{\prime} $ in a general cosmological scenario as shown in (38) \footnote{One can be consistent with the swampland by just taking one of the dS conjectures and so we would only like to discuss the consistency with the original dS conjecture}. This shows that the disagreement found in \cite{kinney2019zoo} between the order of the dS conjecture parameter from the string theoretic motivations and that allowed by the observational data would not arise in a  non-standard cosmological scenario.To illustrate the work we have developed so far, we can take the example of a Chern-Simons cosmology for which the Friedmann equation takes the form (8) \cite{gomez2011standard,del2012approach}. Taking the same ansatz for the Hubble parameter as in \cite{del2012approach}, $ H(\phi) = H_{o} (1 + \beta \phi)$ (where $\beta$ is an arbitrary constant), it is easy to check that the e-folding number bound (32) translates to the requirement \begin{equation}
\alpha H^2 >> 0 
\end{equation}
which would mean that the requirement is for $\alpha$ to be positive as $H^2$ would be positive anyways. Note that such form of Chern-Simons inflation with positive $\alpha$ can easily be consistent with observational data, see \cite{gomez2011standard,del2012approach}. Furthermore, the bound (28) translates to a bound on $H^2$  \begin{equation}
H^2 > \frac{1}{\alpha} (1 - \frac{1}{\beta})
\end{equation}
which can again be satisfied observationally as well. Note also that the bound (40) would also satisfy the $\epsilon$ parameter requirement (35), as (35) is a requirement bore out from the $\epsilon$ parameter issue itself ( as was clear in the initial paper which showed the disagreements \cite{kinney2019zoo}) and is qualitatively the same as the requirements for that and with the $\epsilon$ parameter being appropriately small, the issue with the scalar spectral index also goes away. One can work out more such examples in even more exotic cosmologies but this relatively simple example shows how easily one can check whether inflation in some given cosmology is swampland consistent or not with the approach we have developed here.

\section{Concluding remarks and discussion }
In conclusion, we have shown using quite a general treatment that single field inflationary models can indeed be compatible with the dS and distance conjectures in a wide class of cosmologies where the scalar field follows the usual Klein Gordon Form. We began by discussing in brief about the swampland conjectures and their implications on inflationary cosmology, after which we briefly discussed the exact solution approach to inflation. Then using that approach , we showed systematically how single field inflation in non-standard cosmological scenarios can bypass the problems it faces in GR based cosmology. We showed how the $\epsilon$ parameter for inflation can still be appropriately small for inflation to occur and how the e-fold number can still be up to the scales required by the latest observational data \cite{akrami2018planck}. We then showed that both the observational data for inflation and string theoretic definitions of the c and $ c^{\prime} $ parameters of the dS conjectures and it's refinement , respectively, can agree for single field inflation in non-standard cosmological scenarios. Hence, in essence, we have shown that single field inflation is still very much compatible with the swampland conjectures in a wide class of non-standard cosmological scenarios. Although there has been a lot of work in recent years which have highlighted that non-GR based cosmological scenarios can potentially be better paradigms for single field inflation than their GR counterparts with regards to the swampland, the novelty of our current work lies in the generality of its approach to these issues. While the literature currently on the swampland and single field inflation has analysis based in a specific cosmological background, and then even in only specific potentials in those backgrounds, our approach is not limited to any specific cosmology or potential. It is applicable for all the cosmologies in the scope of our modified friedmann equation and hence shows the compatibility of certain unexplored inflationary regimes with regards to the swampland with conjectures, like inflation in a Tsallis-entropy modified cosmology \cite{sheykhi2018modified} (whilst currently single field inflation in any kind of entropy modifiied cosmologies haven't been studied with regards to the swampland). Hence, the current body of work here is the most general treatment of the swampland issues of single field inflation till now, to the best of our knowledge.
\\
One crucial point which we would like to elaborate here is that in obtaining the equation of the inflaton field we have assumed that the matter, specified by the inflaton
scalar field, enters into the action Lagrangian in such a way that its variation in a Friedmann-Robertson-Walker-Lemaitre background metric leads to the Klein-Gordon equation, expressed by (11). Therefore our method is only applicable to theories where the background metric alongside the perturbations, are not modified.This means that Horava-Lifshitz theories of gravity \cite{mukohyama2010hovrava} or theories of similar plight are beyond the scope of our approach. Hence, we cannot comment on the compatibility of inflationary regimes in such theories with the swampland conjectures. Also, we would also like to clearly state that we do not wish to say that $ L(R) $ theories of gravity which are in the scope of our Friedmann equation (like eq. 9)  are in any way consistent with the swampland conjectures in the way described above, as they are conformally equivalent to a GR based cosmology for an appropriate potential. 

\section*{Acknowledgements}
The author of this paper would like to tremendously thank Sunny Vagnozzi for insightful comments on the paper and the author cannot express enough gratitude for this help. The author would also like to thank the reviewer for their invaluable comments on the manuscript.

\bibliographystyle{unsrt}       % APS-like style for physics
\bibliography{SHREENATHJI.bib}   % name your BibTeX data base

\end{document}